\documentclass[aps,prb,twocolumn,showpacs]{revtex4}

\usepackage{amsmath}
\usepackage{color}
\usepackage{amssymb}

\usepackage{graphicx}

\begin{document}

\title{Superconducting phase transition of Sr$_2$RuO$_4$ in magnetic field}

\author{V.P.Mineev}
\affiliation{Commissariat a l'Energie Atomique, INAC / SPSMS, 38054 Grenoble, France}

\begin{abstract}

 The superconducting state formed due to direct intra-orbital pairing in tetragonal multi-band superconductor Sr$_2$RuO$_4$
 has different properties than the state formed due to intra-band pairing. 
 In particular, the theory operating with   direct intra-orbital pairing successfully explains the  Kerr rotation of reflected light polarization observed several years ago. Here we apply  intra-orbital approach to the problem of Ginzburg-Landau description of  basal plane upper critical field  in this material.
It is shown that  typical for two component superconducting state    additional  phase transition in the vortex state  at $H<H_{c2}$ for all four crystallographic directions of magnetic field in the basal plane and the   basal plane upper critical field anisotropy  still are inevitable properties even in case of direct intra-orbital pairing. 

\end{abstract}

\date{\today}
\maketitle

\section{Introduction}

During about two decades the tetragonal metal   Sr$_2$RuO$_4$ attracts a lot of attention (for the reviews see 
 \cite{Mack,Maeno,Kallin}). 
  In particular, the measurements of the finite Kerr rotation \cite{Xia} in the superconducting phase of this material causes a great interest as a decisive proof for the time reversal symmetry breaking, that is ferromagnetism, spontaneously arising in superconducting  state. 
  A superconducting state  possessing  spontaneous magnetization is described by multicomponent order parameter \cite{Min}. In a tetragonal crystal the  superconducting states with two-component order parameters $(\eta_x,\eta_y)$ corresponding either  to singlet or to triplet pairing are admissible. In application to Sr$_2$RuO$_4$ the triplet pairing state with  time reversal symmetry breaking form of the order parameter 
$(\eta_x,\eta_y)\propto(1,i)$ has been proposed first in the paper \cite{Rice}.

The specific properties for the superconducting state with two-component order parameter  in a tetragonal crystal under magnetic field in basal plane are (i)  the existence of an additional 
phase transition in the vortex state at $H<H_{c2}$ for all four crystallographic directions of magnetic field 
\cite{Agterberg}, and (ii) the anisotropy of the upper critical field \cite{Gor,Bur}. 
Both of these properties should manifest themselves 
starting 
from the Ginzburg-Landau temperature region $T\approx  T_c$ but till now there is no experimental evidence for that. The in-plane  anisotropy of the upper critical field has been observed only at low temperatures \cite{Mao} where it  is quite well known phenomenon for any type of superconductivity
originating from the Fermi surface anisotropy.

Theoretically in application to Sr$_2$RuO$_4$ the phase transition splitting  and the upper critical field anisotropy have been investigated by Agterberg and  co-workers. \cite{Agterberg,Kaur}  They have found that 
one particular choice of  the basis functions of two-dimensional irreducible representation for a tetragonal
point group symmetry  is appropriate for decreasing of basal plane upper critical field anisotropy but at the same time the considerable phase transition splitting occurs. 
Vice versa, another particular choice of the basis functions almost eliminates the phase transition splitting for the  particular field directions but keeps the basal plane upper critical field anisotropy.
Thus the basal plane upper critical field properties look as incompatible with multicomponent order parameter structure dictated by the  experimental observations manifesting the spontaneous time-reversal breaking.

The  theoretical treatment \cite{Kaur}  of $H_{c2}$ problem have been undertaken for the two component
superconducting state in a single band superconductor. In Sr$_2$RuO$_4$ we deal with three bands of charge carriers \cite{Mack}. The
 investigation performed by the present author\cite{Mineev2008} has demonstrated that problems with phase transition splitting and basal plane anisotropy still exist even in multi-band case.

A new opportunity to resolve this inconsistency between the theory and experiments was appeared in relation with  new approach to multiband superconductivity in strontium ruthenate   developed in attempts to explain  the Kerr effect observations \cite{Xia}.
The  Kerr rotation of polarization of light  reflected from the surface of a ferromagnet is expressed through the anomalous Hall conductivity.
 However, there was clearly demonstrated that the Hall conductivity in a single band translationally invariant chiral superconductor is equal to zero.
 \cite{Read,Roy}  In view of this general statement several theoretical explanations for the Kerr effect in clean Sr$_2$RuO$_4$ were proved to be incorrect. Then there were proposed explanations based on skew impurity scattering \cite{Goryo,Lutchin}, such that the Kerr effect has obtained sort of  {\it extrinsic} explanation.
Valuable theoretically these approaches  are seemed to be inappropriate to quite clean strontium ruthenate superconductor.

Recently two groups \cite{Wysokinski,Taylor} showed that in some particular multiorbital superconducting models an {\it intrinsic } anomalous Hall  conductivity does not vanish. Soon after that these results were criticized by the present author \cite{Mineev}, who argued,
on the basis of symmetry considerations and traditional approach to the description of multi-band superconductivity \cite{Suhl},
 that the intrinsic Kerr effect has to vanish even in  a multi-band case. The criticism was mostly addressed to the paper \cite{Taylor}  containing the proper analytic calculation of the Hall conductivity but at the same time based on the model which does not possess general symmetry properties
of a superconducting state in crystal with tetragonal symmetry. However, the following research \cite{Gradhand} has demonstrated that  working  with direct intra-orbital pairing one can  prove the existence of anomalous Hall conductivity in multi-band chiral superconductor.

So,  in frame of direct  intra-orbital pairing approach one can describe  some physical properties which don't give in to an explanation in terms of traditional description of multi-band superconductivity.  Here we  develop the Ginzburg-Landau  theory of multi-band unconventional superconductivity based entirely on the intra-orbital pairing.
 It is shown  that typical for  two component superconducting state phase transition splitting under basal plane magnetic field in all four crystallographic directions 
is  obligatory even  in case of direct intra-orbital pairing. The basal plane upper critical field anisotropy takes place as well.

\section{Intra-band  versus intra-orbital pairing approaches to multi-band superconductivity}

The conducting bands in Sr$_2$RuO$_4$  are formed by three Ru-$4d$ orbitals $d_{xz},d_{yz},d_{xy},$ denoted as $a,b,c$ respectively. \
 The dispersion $\varepsilon_c({\bf k})$ of $c$  band  has  tetragonal symmetry, but the symmetry of dispersion of $a$ and $b$ bands $\varepsilon_a({\bf k})$ and $\varepsilon_b({\bf k})$ is orthorhombic. The full tetragonal symmetry of real normal state  
 is recreated after including the interband $\varepsilon_{ab}({\bf k}),\varepsilon_{ac}({\bf k}),\varepsilon_{bc}({\bf k})$ coupling matrix elements and diagonalization
of total Hamiltonian. As result, from initial $\varepsilon_a({\bf k}),\varepsilon_b({\bf k}),\varepsilon_c({\bf k})$ bands dispersion we  come to dispersion laws $E_\alpha({\bf k}),E_\beta({\bf k}),E_\gamma({\bf k})$  of Sr$_2$RuO$_4$ $\alpha,\beta,\gamma$  bands  \cite{Mack} possessing full tetragonal symmetry.

The regular procedure accepted in multi-band superconductivity theory is the following. First, one  must perform  the normal state band Hamiltonian diagonalisation and then introduce pairing between the electrons filling the states in the bands
with full tetragonal symmetry.  It is correct not only from the symmetry point of view but also because usually the normal state band splitting is much larger than the thickness of the layer near the Fermi surface where a pairing interaction is effective. 
Applying then the Bogolubov transformation  to the normal state band Hamiltonians for bands $\alpha,\beta,\gamma$, intraband pairing and  interband pair scattering terms we come to mathematical description of {\it multiband} superconducting state. \cite{Suhl}
Following this procedure and then calculating current response one can prove that Hall conductivity in a multi-band superconducting state vanishes completely.

Another approach is to  introduce direct
pairing   between the electrons filling the initial orbital bands, and then,  taking into account inter-band normal state hoping amplitudes like $\varepsilon_{bc}$, to make diagonalisation of total Hamiltonian 
including all the superconducting and normal  parts.
So, the diagonalisation of normal orbital parts and the pairing terms is produced simultaneously. 
This approach has a sense when at any band splitting still  there is  an attraction between the electrons with opposite momenta filling initial orbital states. The procedure is resulted in formation of 
{\it multi-orbital} superconducting state possessing  nonzero Hall conductivity.\cite{Gradhand}

In what follows we compare  the  upper critical field problem description in  $\alpha,\beta,\gamma$ band representation and in $a,b,c$ band representation.
 We shall discuss triplet unitary  superconducting state
  having only one  spin component $|\uparrow\downarrow\rangle+|\downarrow\uparrow\rangle$ corresponding to the equal spin pairing with spins perpendicular to the spin quantization axis   chosen parallel to the tetragonal axis $\hat z$. 
  
  For mathematical simplicity as in Ref.11 we  limit ourself by the description of   two bands  $\alpha,\beta$ and  $a,b$ situation. One can demonstrate that addition of the third band making the treatment much more cumbersome does not change
results qualitatively. Also we shall ignore inter-orbital spin-orbital coupling (it can be included
following the paper \cite{Ng}) that 
 introduces nothing important in the description of triplet superconducting state put forward here.

Following Ref.18,21  for the description of the normal state electronic band structure we consider:
\begin{eqnarray}
&\varepsilon_{a{\bf k}}+\mu=-2t\cos k_x-2t^\perp\cos k_y,\nonumber\\
&\varepsilon_{b{\bf k}}+\mu=-2t\cos k_y-2t^\perp\cos k_x,
\end{eqnarray}
\begin{equation}
\varepsilon_{ab{\bf k}}=-2t^{\prime\prime}\sin k_x\sin k_y.
\end{equation}
Corresponding $\alpha,\beta$ bands dispersion laws are
\begin{equation}
E_{\alpha,\beta{\bf k}}=\frac{1}{2}(\varepsilon_{a{\bf k}}+\varepsilon_{b{\bf k}})\mp\frac{1}{2}\sqrt{(\varepsilon_{a{\bf k}}-\varepsilon_{b{\bf k}})^2+4\varepsilon_{ab{\bf k}}^2}.
\end{equation}

\subsection{H$_{c2}$ problem  in superconducting state with $\alpha,\beta$ intra-band pairing}

The general form of two component order parameter corresponding to each band is 
\begin{equation}
\Delta_\lambda({\bf k},{\bf q})=\eta_{\lambda x}({\bf q})\varphi_{\lambda x}({\bf k})+\eta_{\lambda y}({\bf q})\varphi_{\lambda y}({\bf k}),
\label{D}
\end{equation}
where $\lambda$  runs the band labels $ \alpha,\beta$,
$(\varphi_{\lambda x},\varphi_{\lambda y})$ are basis functions of two dimensional representation of tetragonal group transforming as $(k_x,k_y)$. In general they are different for the different bands.

The upper critical field is determined as the eigen value of linear equation for the order parameter
\begin{eqnarray}
\Delta_{\lambda}({\bf k},{\bf q})=T\sum_{n}\int\frac{d^3{\bf k}^\prime}{(2\pi)^3}\sum_{\mu} 
V_{\lambda\mu} \left( {\bf k},{\bf k}'\right)\nonumber\\
\times 
G_\mu({\bf k}',\omega_n)G_\mu(-{\bf k}'+{\bf q},-\omega_n)\Delta_{\mu}({\bf k}',{\bf q}). 
\label{gap eq}
\end{eqnarray}
Here
\begin{equation}
V_{\lambda\mu}( {\bf k},{\bf k}')=V_{\lambda\mu}\sum_{i=x,y}\varphi_{\lambda i}({\bf k})\varphi_{\mu i}({\bf k}')
\end{equation}
is the pairing interaction matrix, and
\begin{equation}
G_\mu({\bf k},\omega_n)=\frac{1}{i\omega_n-E_{\mu{\bf k}}}
\end{equation}
is the normal state band $\mu$ Green function.

Performing the Taylor expansion of  equation (\ref{gap eq}) in powers of ${\bf q}$ up to the second order and transforming to the 
coordinate representation, that means simple substitution
\begin{equation}
{\bf q}\to {\bf D}=-i\nabla_{\bf r}+2e{\bf A}({\bf r}),
\end{equation}
we obtain Ginzburg-Landau equations 
\begin{equation}
\eta_{\lambda i}({\bf r})=\sum_{\mu j} V_{\lambda\mu}(L^\mu_{ij}+M^\mu_{ijlm}D_lD_m)\eta_{\mu j}({\bf r}),
\label{eqn}
\end{equation}
where 
\begin{equation}
L^\mu_{ij}
=T\sum_{n}\int\frac{d^3{\bf k}}{(2\pi)^3}~\varphi_{\mu i}({\bf k})\varphi_{\mu j}({\bf k})
G_\mu({\bf k},\omega_n)G_\mu(-{\bf k},-\omega_n),
\end{equation}
\begin{eqnarray}
M^\mu_{ijlm}=
\frac{T}{2}\sum_{n}\int\frac{d^3{\bf k}}{(2\pi)^3}\varphi_{\mu i}({\bf k})\varphi_{\mu j}({\bf k})\\
\times G_\mu({\bf k},\omega_n)\frac{\partial^2G_\mu(-{\bf k},-\omega_n)}{\partial k_l\partial k_m}.
\label{oper}
\end{eqnarray}
In absence of magnetic field ${\bf D}=0$ and taking into account that
\begin{equation}
L^\mu_{ij}=L^\mu_{xx}\delta_{ij}=L^\mu_{yy}\delta_{ij}
\end{equation}
we come to two separate  systems of homogeneous equations 
\begin{equation}
\eta_{\lambda i}({\bf r})=\sum_{\mu} V_{\lambda\mu}L^\mu_{xx}\eta_{\mu i}({\bf r}).
\label{eqn1}
\end{equation}
for $x$ and $y$ components of the order parameter for both bands. Determinant of each of them determines the critical temperature of phase transition. These systems are completely equivalent each other.
Hence, the phase transition to superconducting state occurs at the same critical temperature for all the component of the order parameter in all the bands.  
Simple BCS-like formula for $T_c$  was pointed out in Ref.12 where it was found in assumption that logarithmically divergent terms $L^\mu_{xx}$ have the same energy cutoff in different bands. In general it is not true and an expression for critical temperature is more cumbersome.

Relative value of
 $x$ and $y$ components of the order parameter  is fixed by the nonlinear terms in GL equations. For single band superconductors this problem was solved by Volovik and Gor'kov \cite{Volovik}.
 There was shown that the complex superconducting state arising directly from the normal state by means of phase transition of second order always has form $\vec\eta=\eta(1,i)$. The equality of modulus of $x$ and $y$ components of the order parameter guarantees the minimum of GL free energy.

Unlike single band superconductivity in two band case the complex superconducting state with
order parameter   $\vec\eta_\alpha=(\eta_{\alpha x},\eta_{\alpha y}),~~\vec\eta_\beta=(\eta_{\beta x},\eta_{\beta y})$  in bands $\alpha$ and $\beta$ does not oblige to have equal modulus of x and y components (see Appendix).

In presence of a magnetic field in the basal plane\\ $ {\bf H}=H(\cos\varphi,\sin\varphi,0),$
chosing the vector potential as 
\begin{equation}
{\bf A}=H(0,0,y\cos\varphi-x\sin\varphi),
\label{A}
\end{equation}
such that  
\begin{eqnarray}
&D_x=-i\frac{\partial}{\partial x},~~~~D_y=-i\frac{\partial}{\partial y},\nonumber\\
&D_z=-i\frac{\partial}{\partial z}+2eH(y\cos\varphi-x\sin\varphi)
\label{D}
\end{eqnarray}
the  equations determining the upper critical field  acquire the form
\begin{widetext}
\begin{equation}
\left( \begin{array} {cccc}\eta_{\lambda x}\\
\eta_{\lambda y}
\end{array}\right )=\sum_\mu V_{\lambda\mu}\left( \begin{array} {cccc}
L_{xx}^\mu+M_{xxxx}^\mu D_x^2+M_{xxyy}^\mu D_y^2+M_{xxzz}^{\mu}D_z^2
& 2M_{xyxy}^\mu D_xD_y\\
2M_{xyxy}^\mu D_xD_y&
L_{yy}^\mu+M_{yyyy}^\mu D_y^2+M_{yyxx}^\mu D_x^2+M_{yyzz}^{\mu}D_z^2
\end{array}\right )\left( \begin{array} {cccc}\eta_{\mu x}\\
\eta_{\mu y}
\end{array}\right ).
\label{H}
\end{equation}
\end{widetext}
For arbitrary  field direction in the basal plane 
 the system of equations for the $x$ components of the order parameters is entangled with the system of equation for $y$ components.
At ${\bf H}\parallel \hat x$, the dependence from $x$ coordinate drops out
and the system of equations is split into  independent systems of equations for $x$ and $y$ components of the order parameter.
 \begin{eqnarray}
&\eta_{\lambda x}=\sum_\mu V_{\lambda\mu}(L_{xx}^\mu+
M_{xxyy}^\mu D_y^2+M_{xxzz}^{\mu}D_z^2)\eta_{\mu x},\nonumber\\
&\eta_{\lambda y}=\sum_\mu V_{\lambda\mu}
(L_{yy}^\mu+M_{yyyy}^\mu D_y^2+M_{yyzz}^{\mu}D_z^2)\eta_{\mu y}.
\end{eqnarray}

  Let us assume that the maximum critical field is determined by system of equations for the $y$ components of the order parameter. Its solution is given by functions  independent of $x$ coordinate 
\begin{equation}
\eta_{\lambda y}=\eta_{\lambda y}(y,z), ~~~\lambda=\alpha,\beta.
\label{A}
\end{equation}
Hence, the order parameters in both bands
\begin{equation}
\eta_{\lambda y}(y,z)\varphi_{\lambda y}({\bf k}),~~~\lambda=\alpha,\beta.
\label{B}
\end{equation}
are invariant under reflection $\sigma_x$ about $\hat x$ direction.
At the same time, two component zero field order parameter for both bands 
\begin{equation}
\eta_{\lambda x}\varphi_{\lambda x}({\bf k})+\eta_{\lambda y}\varphi_{\lambda y}({\bf k}), ~~\lambda=\alpha,\beta
\label{C}
\end{equation}
does not possess the $\sigma_x$ symmetry. Hence, exactly as  in single band case \cite{Agterberg}, there must exist a second transition in the finite field $H<H_{c2}$  at which $\eta_{\alpha x}$
and $\eta_{\beta x}$  become nonzero. Similar arguments hold for the field along any of the other three crystallographic directions in the basal plane.  The existence of two transitions for all four crystallographic axes in the basal plane is a consequence of two component structure of the order parameter in each band, which can be  either real or complex.

For the arbitrary direction of magnetic field in the basal plane one can diagonalize the system  (\ref{H}) directly demonstrating the anisotropy of $H_{c2}(\varphi)$. \cite{Mineev2008} However, at arbitrary field direction the order parameter does not obey the symmetry in respect to reflection in the plane perpendicular to field direction, so the second phase transition is not obliged to be present.\cite{fn}

\subsection{H$_{c2}$ problem  in superconducting state with $a,b$ intra-band pairing}

When we deal with direct pairing of electrons filling the states in non hybridized bands $a$ and $b$  the  two component order parameter
corresponding to each band keeps the same form 
\begin{equation}
\Delta_u({\bf k},{\bf q})=\eta_{u x}({\bf q})\varphi_{x}({\bf k})+\eta_{uy}({\bf q})\varphi_{y}({\bf k}),
\end{equation}
where index $u$  runs the band labels $ a,b$,
$(\varphi_{x},\varphi_{y})$ are basis functions of two dimensional representation of tetragonal group transforming as $(k_x,k_y)$. Unlike
situation discussed in previous subsection they are the same for the different bands. 

The linear equation for order parameter
 acquire the following form
\begin{widetext}
\begin{eqnarray}
\Delta_{u}({\bf k},{\bf q})=T\sum_{n}\int\frac{d^3{\bf k}^\prime}{(2\pi)^3}\left [\sum_{w} V_{uw} ( {\bf k},{\bf k}'
G_w({\bf k}',\omega_n)G_w(-{\bf k}'+{\bf q},-\omega_n)\Delta_{w}({\bf k}',{\bf q})\right.\nonumber\\
\left.+\sum_{\{vw\}} V_{uv} \left( {\bf k},{\bf k}'\right)
G_{vw}({\bf k}^\prime,\omega_n)G_{vw}(-{\bf k}'+{\bf q},-\omega_n)\Delta_{w}({\bf k}',{\bf q})\right ], 
\label{gap eq4}
\end{eqnarray}
\end{widetext}
where sign $\sum_{\{vw\}}$ denotes summation over $v$ and $w$ at ${v\ne w}$.
Here
\begin{equation}
V_{uw}( {\bf k},{\bf k}')=V_{uw}\sum_{i=x,y}\varphi_{i}({\bf k})\varphi_{i}({\bf k}')
\end{equation}
is the pairing interaction matrix with properties
\begin{equation}
V_{aa}=V_{bb}, ~~~V_{ab}=V_{ba},
\end{equation}
 and
\begin{equation}
G_a({\bf k},\omega_n)=\frac{i\omega_n-\varepsilon_{b{\bf k}}}{(i\omega_n-E_{\alpha{\bf k}})(i\omega_n-E_{\beta{\bf k}})}
\end{equation}
\begin{equation}
G_b({\bf k},\omega_n)=\frac{i\omega_n-\varepsilon_{a{\bf k}}}{(i\omega_n-E_{\alpha{\bf k}})(i\omega_n-E_{\beta{\bf k}})}
\end{equation}
\begin{equation}
G_{ab}({\bf k},\omega_n)=G_{ba}({\bf k},\omega_n)=\frac{\varepsilon_{ab{\bf k}}}{(i\omega_n-E_{\alpha{\bf k}})(i\omega_n-E_{\beta{\bf k}})}
\end{equation}
are the normal state bands $a,b$ and the interbands $ ab, ba$ Green functions.

Transforming order parameter equation to the coordinate representation and keeping only terms up to the second order in gradients we obtain Ginzburg-Landau equations 
\begin{eqnarray}
\eta_{u i}({\bf r})=\sum_{w j} V_{uw}(L^w_{ij}+M^w _{ijlm}D_lD_m)\eta_{w j}({\bf r})\nonumber\\
+\sum_{\{vw\} j} V_{uv}(L^{vw}_{ij}+M^{vw} _{ijlm}D_lD_m)\eta_{w j}({\bf r}),
\label{eqn5}
\end{eqnarray}
where 
\begin{equation}
L^w_{ij}
=T\sum_{n}\int d^3{\bf k}~\varphi_{i}({\bf k})\varphi_{ j}({\bf k})
G_w({\bf k},\omega_n)G_w(-{\bf k},-\omega_n),
\end{equation}
\begin{eqnarray}
M^w_{ijlm}=
\frac{T}{2}\sum_{n}\int d^3{\bf k}~\varphi_{ i}({\bf k})\varphi_{ j}({\bf k})\\
\times G_w({\bf k},\omega_n)\frac{\partial^2G_w(-{\bf k},-\omega_n)}{\partial k_l\partial k_m},
\label{oper1}
\end{eqnarray}
\begin{equation}
L^{vw}_{ij}
=T\sum_{n}\int d^3{\bf k}~\varphi_{i}({\bf k})\varphi_{ j}({\bf k})
G_{vw}({\bf k},\omega_n)G_{vw}(-{\bf k},-\omega_n),
\end{equation}
\begin{eqnarray}
M^{vw}_{ijlm}=
\frac{T}{2}\sum_{n}\int d^3{\bf k}~\varphi_{ i}({\bf k})\varphi_{ j}({\bf k})\\
\times G_{vw}({\bf k},\omega_n)\frac{\partial^2G_{vw}(-{\bf k},-\omega_n)}{\partial k_l\partial k_m},
\label{oper1}
\end{eqnarray}

In absence of magnetic field we have  two separate systems of homogeneous equations 
\begin{equation}
\eta_{u i}({\bf r})=\sum_{w j} V_{uw}L^w_{ij}\eta_{w j}({\bf r})
+\sum_{\{vw\} j} V_{uv}L^{vw}_{ij}\eta_{w j}({\bf r})
\end{equation}
for $x$ and $y$ components of the order parameter for both bands. Determinant of each of them determines the critical temperature of phase transition.
Using explicit expressions for the Green functions, bands dispersion laws, and pairing interaction properties one can be convinced in following
symmetry relations
\begin{eqnarray}
&L_{xx}^a=L_{yy}^b,~~~L_{xx}^b=L_{yy}^a,~~~L_{xy}^{a,b}=L_{yx}^{a,b}=0\nonumber\\
&L_{xx}^{ab}=L_{yy}^{ab}=L_{xx}^{ba}=L_{yy}^{ba},\nonumber\\
&L_{xy}^{ab}=L_{yx}^{ab}=L_{xy}^{ba}=L_{yx}^{ba}=0.~~
\end{eqnarray}
Then, it is easy to see that system of equations for $(\eta_{ax},\eta_{bx})$ completely coincides with 
system of equations for $(\eta_{by},\eta_{ay})$.
As result, the phase transition to superconducting state occurs at the same critical temperature for all the component of the order parameter in all the bands. 

Relative value of x and y components of the order parameter is fixed by the nonlinear terms in GL equations. Unlike single band superconductivity in two band case the complex superconducting state with
order parameters  $\vec\eta_a=(\eta_{ax},\eta_{ay}),~~\vec\eta_b=(\eta_{bx},\eta_{by})$  in bands $a$ and $b$ does not oblige to have equal modulus of x and y components (see Appendix). 
For example, authors of paper \cite{Gradhand} consider the state with order parameters with following relationship between components
\begin{equation}
\eta_{ay}=i\eta_{bx},~~~\eta_{by}=i\eta_{ax}. 
\label{antianti}
\end{equation}

In presence of magnetic field gradient terms mix the systems of equations for $x$ and $y$ components of the order parameter
\begin{widetext}
\begin{eqnarray}
\left( \begin{array} {cccc}\eta_{u x}\\
\eta_{uy}
\end{array}\right )=\sum_w V_{uw}\left( \begin{array} {cccc}
L_{xx}^w+M_{xxxx}^w D_x^2+M_{xxyy}^w D_y^2+M_{xxzz}^wD_z^2
& 2M_{xyxy}^w D_xD_y\\
2M_{xyxy}^w D_xD_y&
L_{yy}^w+M_{yyyy}^w D_y^2+M_{yyxx}^w D_x^2+M_{yyzz}^{\mu}D_z^2
\end{array}\right )\left( \begin{array} {cccc}\eta_{w x}\\
\eta_{w y}\end{array}\right )\nonumber\\
+
\sum_{\{vw\}} V_{uv}\left( \begin{array} {cccc}
L_{xx}^{vw}+M_{xxxx}^{vw} D_x^2+M_{xxyy}^{vw} D_y^2+M_{xxzz}^{vw}D_z^2
& 2M_{xyxy}^{vw} D_xD_y\\
2M_{xyxy}^{vw} D_xD_y&
L_{yy}^{vw}+M_{yyyy}^{vw} D_y^2+M_{yyxx}^{vw} D_x^2+M_{yyzz}^{vw}D_z^2
\end{array}\right )\left( \begin{array} {cccc}\eta_{w x}\\
\eta_{w y}
\end{array}\right ).
\label{H3}
\end{eqnarray}

At ${\bf H}\parallel \hat x$ the dependence of $x$ coordinate drops out
and the system of  equations is split into two independent systems of equations for $x$ and $y$ components of the order parameter
\begin{eqnarray}
&\eta_{u x}=\sum_w V_{uw}(L_{xx}^w+M_{xxyy}^w D_y^2+M_{xxzz}^wD_z^2)\eta_{w x}+
\sum_{\{vw\}} V_{uv}(L_{xx}^{vw}+M_{xxyy}^{vw} D_y^2+M_{xxzz}^{vw}D_z^2)\eta_{w x},\nonumber\\
&\eta_{uy}=\sum_w V_{uw}(L_{yy}^w+M_{yyyy}^w D_y^2+M_{yyzz}^{\mu}D_z^2)\eta_{w y}+
\sum_{\{vw\}} V_{uv}(L_{yy}^{vw}+M_{yyyy}^{vw} D_y^2+M_{yyzz}^{vw}D_z^2)\eta_{w y}.
\label{H4}
\end{eqnarray}
\end{widetext}
Argumentation  about the symmetry of the order parameter given at the end of previous section  is applicable here as well.
Solution of Eq. (\ref{H4})  for $y$ order parameter components is given by functions  independent of $x$ coordinate 
\begin{equation}
\eta_{u y}=\eta_{uy}(y,z), ~~~u=a,b.
\label{A'}
\end{equation}
Hence, the order parameters in both bands
\begin{equation}
\eta_{uy}(y,z)\varphi_{uy}({\bf k}),~~~u=a,b.
\label{B'}
\end{equation}
are invariant under reflection $\sigma_x$ about $\hat x$ direction.
At the same time, two component zero field order parameter for both bands 
\begin{equation}
\eta_{ux}\varphi_{ux}({\bf k})+\eta_{uy}\varphi_{uy}({\bf k}), ~~~u=a,b
\label{C'}
\end{equation}
does not possess the $\sigma_x$ symmetry.

Hence, even in case of direct intra-orbital pairing, there must exist a second transition in the finite field $H<H_{c2}$ at which $\eta_{ax}$
and $\eta_{bx}$  become nonzero. Similar arguments hold for the field along any of the other three crystallographic directions in the basal plane.  The existence of two transitions for all four crystallographic axes in the basal plane is a consequence of two component structure of the order parameter in each band, which can be  either real or complex.  At arbitrary field direction the order parameter does not obey the symmetry in respect to reflection in the plane perpendicular to field direction, so the second phase transition is not obliged to be present.

For the arbitrary direction of magnetic field in the basal plane one cannot diagonalize the system  (\ref{H4}) but obviously the anisotropy of $H_{c2}(\varphi)$ still takes place.

\section{Conclusion}

We have demonstrated that  there is an additional  phase transition in the vortex state at $H<H_{c2}$ for four $\pm\hat x,\pm \hat y$ magnetic field  directions  in the basal plane
in 
a multicomponent tetragonal superconductor.
Being independent of intra-band or intra-orbital type of pairing 
this quality and also the upper critical field anisotropy in the basal plane are the inherent properties of multicomponent superconducting state with tetragonal symmetry.

\section*{Acknowledgements}
I am indebted to M. E. Zhitomirsky for useful discussions.

\appendix

\section{Two component two-band superconductivity}
\subsection{One band superconductivity}
The GL free energy functional for the two component superconducting state with order parameter
\begin{equation}
(\vec\eta\cdot\vec\varphi({\bf k}))=\eta_x\varphi_x({\bf k})+\eta_y\varphi_y({\bf k})
\end{equation}
 in tetragonal superconductor is
\begin{equation}
F=\alpha_0(T-Tc){\vec\eta}{\vec \eta}^* +\beta_1({\vec\eta}{\vec \eta}^*)^2+\beta_2|{\vec\eta}{\vec \eta}|^2+
\beta_3(|\eta_{x}|^4+|\eta_{y}|^4)
\label{L}
\end{equation}
Here $(\varphi_{ x},\varphi_{ y})$ are basis functions of two dimensional representation of tetragonal group transforming as $(k_x,k_y)$. 

 There was shown  \cite{Volovik} that at $\beta_2>0$, and $\beta_3>-2\beta_2$ the state
with two component complex order parameter $\vec\eta=\eta(1,i)$ arising  by the second order phase transition directly from the normal state  realizes the absolute minimum of the free energy density (\ref{L}). 
This state belongs to the superconducting class
\begin{equation}
D_4(E)=\left\{ \exp\left (\frac{i\pi n}{2}\right ) C_n, \exp\left (-\frac{i\pi n}{2}\right ) RU_n\right\},
\label{D4}
\end{equation}
which is the symmetry group of the order parameter.
Here $n=0,1,2,3$, and $C_n$ are rotation around $\hat z$-axis on angles $\pi n/2$, $U_n$ are rotations on angle $\pi$ around
axes $\hat x,  \hat x+\hat y, \hat y,-\hat x+\hat y$ correspondingly.
The group $D_4(E)$ is the subgroup of maximal symmetry  of the group of symmetry G of the normal state given by direct product of tetragonal point group, group of gauge transformations and the group of time inversion 
\begin{equation}
G=D_{4h}\times U(1)\times R.
\label{G}
\end{equation}
The symmetry $D_4(E)$ of the order parameter guarantees invariance of GL free energy (\ref{L})  in respect to all transformations  of normal state symmetry group G.  
 
The superconducting state  with order parameter 
\begin{equation}
\vec\eta=\eta(1,ir),~
r\ne 1,
\end{equation} 
where $r$ is real number, can arise from the superconducting state with tetragonal symmetry $D_4(E)$ at some lower temperature $\tilde T_c<T_c$.
The corresponding free energy term responsible for this phase transition
\begin{equation}
\tilde\alpha_0(T-\tilde T_c)(\eta_x+i\eta_y)(\eta_x^*-i\eta_y^*)
\end{equation}
is not invariant in respect of symmetry group G of the normal state. But it is invariant in respect to all transformations  of the group $D_4(E)$ of the superconducting state  with order parameter $\vec\eta=\eta(1, i)$.
The symmetry of  state $\vec\eta=\eta(1,ir),~
r\ne 1$  is given by orthorhombic group
\begin{equation}
D_2(E)=\left\{ \exp\left (\frac{i\pi n}{2}\right ) C_n, \exp\left (-\frac{i\pi n}{2}\right ) RU_n\right\}
\end{equation}
Here, unlike to Eq. (\ref{D4}) index $n=0,2$ only. The group $D_2(E)$ is a subgroup of the group $D_4(E)$.

\subsection{Two band superconductivity}

 Unlike single band superconductivity a two band state with band order parameters such that 
\begin{equation}
i\eta_{ux}=\eta_{uy},~~~u=a,b
\label{s}
\end{equation}
 is not in general the absolute minimum of free energy.
It is because along with the invariant $\beta_2^{\prime}(|{\vec\eta_a}{\vec \eta_a}|^2+|{\vec\eta_b}{\vec \eta_b}|^2)$ vanishing at fulfilment (\ref{s}) decreasing free energy at $\beta_2^{\prime}>0$ 
there are several mixing terms in free energy expansion. For instance, the term  ${\tilde\beta}({\vec \eta_a}{\vec \eta_b}^*)({\vec \eta_a}^*{\vec \eta_b})$ obviously works to decrease free energy at ${\tilde\beta}>0$  when modulus of $x$ and $y$ components of the order parameter are quite different.

Authors of paper \cite{Gradhand} consider the state with order parameters with following relationship between components
\begin{equation}
\eta_{ay}=i\eta_{bx},~~~\eta_{by}=i\eta_{ax}. 
\label{p}
\end{equation}
This state with symmetry $D_2(E)$ can arise   directly from the normal state  by the second order phase transition.


\begin{thebibliography}{20}
\bibitem{fn} Note that unlike to the tetragonal crystal for the hexagonal superconductor the
additional phase transition must exist at arbitrary basal plane field ddirection.
\end{thebibliography}
\end{document}